\documentclass{llncs}
\usepackage{amsfonts}

\begin{document}

\newcommand{\dima}[1]{{{#1}}}                
\newcommand{\mik}[1]{{{#1}}}           
\newcommand{\cris}[1]{{{#1}}}           
\newcommand{\poly}{{\rm poly}}
\newcommand{\eps}{\epsilon}
\newcommand{\tQ}{\tilde{Q}}
\newcommand{\tP}{\tilde{P}}
\newcommand{\tE}{\tilde{E}}
\newcommand{\tlambda}{\tilde{\lambda}}
\newcommand{\ttau}{\tilde{\tau}}
\newcommand{\Z}{{\mathbb Z}}
\newcommand{\D}{\Delta}
\newcommand{\proofend}{\hspace*{\fill}\mbox{$\Box$}}
\newcommand{\inv}[1]{{1\over #1}}
\newcommand{\mult}{{\rm mult}}

\newcommand{\remove}[1]{}

\def\e{\epsilon}

\title{Rapid Mixing for Lattice Colorings with Fewer Colors}


\author{Dimitris Achlioptas\inst{1} \and Mike Molloy\inst{2}
 \and Cristopher Moore\inst{3} \and Frank Van Bussel\inst{4}}

\institute{Microsoft Research {\tt optas@microsoft.com}
\and Dept of Computer Science, University of Toronto, and Microsoft
Research \thanks{Some of this work was done while visiting the Fields
Institute. Supported by NSERC and a Sloan Research Fellowship}
{\tt molloy@cs.toronto.edu}
\and Computer Science Department, University of New Mexico
\thanks{Supported by NSF grants PHY-0200909, CCR-0220070, and EIA-0218563} {\tt moore@santafe.edu}
\and Dept of Computer Science, University of Toronto
{\tt fvb@cs.toronto.edu}}

\maketitle

\remove{
\begin{center}
{\bf Contact author:} M. Molloy, Dept of Computer Science, University of Toronto,
10 Kings College Rd, Toronto, ON M5S 3G1, Canada.
\end{center}
}

\begin{abstract}
We provide an optimally mixing Markov chain for 6-colorings of the
square lattice \cris{on rectangular regions with free, fixed, or toroidal boundary conditions}.  
This implies that the uniform distribution on
the set of such colorings has strong spatial mixing, 
\cris{so that the 6-state Potts antiferromagnet has a finite correlation length 
and a unique Gibbs measure at zero temperature.}
Four and five are now the only remaining values of $q$ for which it is not known 
whether there exists a rapidly mixing Markov chain for $q$-colorings of the
square lattice.
\end{abstract}

\section{Introduction}

Sampling and counting \dima{graph colorings} is a fundamental
problem in computer science and discrete mathematics.  \cris{It is}
also of fundamental interest in statistical physics: graph colorings correspond to the
zero-temperature case of the {\em antiferromagnetic Potts model},
a model of magnetism on which physicists have performed extensive
numerical experiments (see for instance~\cite{ferreirasokal,vortex,moorenewman}). 
\remove{
\mik{Unless specified otherwise,
we will always be referring to the zero-temperature case of the antiferromagnetic
Potts model throughout this paper.}
}

Physicists wish to estimate quantities such as spatial
correlations and magnetization, and to do this they 
sample random states using \dima{Markov chains}.  \mik{This is a general technique
whereby one starts with an arbitrary state, and then repeatedly modifies it
using a random rule.  For the zero-temperature Potts model, two standard such
rules described below are Glauber dynamics and Kempe chain flips, also known as the 
Wang-Swendsen-Koteck{\'y} algorithm.}  \cris{While these algorithms often appear to 
work well in practice, we would like to know that their {\em mixing time}---i.e., the time 
it takes them to achieve a nearly-uniform distribution on the set of states---is polynomially 
bounded as a function of the size of the lattice.  Establishing this rigorously has been a major project in mathematical physics and theoretical computer science; see for example~\cite{bdg,j,martinelli,salassokal,ev}.}

Moreover, {\em optimal temporal mixing,} i.e., a mixing time of
$O(n \log n)$, is deeply related to the physical properties of the
system~\cite{dsvw}.  In particular, under certain conditions on the Markov chain, 
it implies {\em spatial mixing}, i.e., the exponential decay of correlations, 
and thus the existence of a finite correlation length and the uniqueness of the
Gibbs measure.  Therefore, optimal mixing of natural Markov chains
for $q$-colorings of the square lattice is considered a major open problem
in physics (see e.g.~\cite{sokalunsolved}). Physicists have
conjectured~\cite{ferreirasokal,sokalunsolved} that the $q$-state
Potts model has spatial mixing for $q \ge 4$, 
\cris{even at zero temperature}.  As we discuss below, 
previous results~\cite{j,salassokal,bdg,bdgj} have established this rigorously 
for $q \ge 7$.
\remove{
\mik{The case $q\ge9$ follows
from the well-known results of Jerrum~\cite{j} and of Salas and Sokal~\cite{salassokal}
which establish optimal mixing for any graph of maximal degree $\D$ when $q\ge 2\D+1$.
The case $q\ge7$ has been established rigorously by Bubley, Dyer and
Greenhill~\cite{bdg} \mik{(see also~\cite{bdgj})} who showed that all triangle-free
graphs with maximum degree 4}, which includes the square lattice, have optimal mixing.
}

Our main result is that the square lattice has \cris{strong spatial} mixing for
$q=6$, \cris{and in particular that a natural Markov chain has optimal temporal mixing.  
We consider the so-called} {\em block heat-bath dynamics}, which we call
$M(i,j)$. At each step, \cris{for fixed integers $i$ and $j$,} we choose an $i \times j$ block $S$ of
\cris{the lattice} $G$, uniformly at random from among all such blocks 
\cris{contained in $G$.}
\remove{(i.e., its upper-left vertex is chosen uniformly from the vertices of $G$).}
Let $C$ be the set of $q$-colorings of $S$ which are consistent
with the coloring of $G \setminus S$.  We choose a uniformly
random coloring $c \in C$ and recolor $S$ with $c$. 
Our main theorem is:
\begin{theorem}\label{t1}
$M(2,3)$ on 6-colorings of the square lattice mixes in $O(n \log n)$
time.
\end{theorem}

We prove Theorem~\ref{t1} for 
\cris{finite rectangular regions with free, fixed, or toroidal boundary conditions.}
\dima{Our method is similar to that of~\cite{bdg} in that it
consists of a computer-assisted proof of the existence of a path
coupling. At the same time, we exploit the specific geometry of
the square lattice to} consider a greater variety of neighborhoods.
Moreover, the calculations \dima{necessary to find a good coupling
in our setting are far more complicated than those in~\cite{bdg}
and require several new ideas to become computationally
tractable.}

Using the comparison method of Diaconis and
Saloff-Coste~\cite{ds-c,randalltetali}, Theorem \ref{t1}
implies that the Glauber and Kempe chain Markov chains also mix in
polynomial time:
\begin{theorem}\label{t2}
The \cris{Glauber and Kempe chain Markov chains} on 6-colorings
of the square lattice mix in $O(n^2 \log n)$ time.
\end{theorem}
Like Theorem~\ref{t1}, this result holds on \cris{finite rectangular regions with 
free, fixed, or toroidal boundary conditions.}

\cris{To discuss spatial mixing, suppose we have} 
a finite region $V$ and two colorings $C, C'$
of its boundary that differ at a single vertex $v$, and a subregion
$U \subseteq V$ such that the distance from $v$ to the nearest
point $u \in U$ is $\ell$. Let $\mu$ and $\mu'$ denote the
probability distributions on colorings of $U$, given the uniform
distribution on colorings of $V$ conditioned on $C$ and $C'$
respectively.   
Then we define spatial mixing as follows: 

\begin{definition}[\cite{dsvw}] We say that $q$-colorings have 
{\em strong spatial mixing} if there are constants $\alpha, \beta > 0$ such that 
\cris{the total variation distance between $\mu$ and $\mu'$ obeys}
$\| \mu - \mu' \| \le \beta |U| \exp(-\alpha \ell)$. 
\end{definition}

In other words, \cris{strong spatial mixing means that} 
conditioning \cris{a uniformly random coloring of the lattice} 
on the event that particular colors appear on vertices
far away from $v$ \dima{has an exponentially small effect on the
conditional distribution of the color of $v$.}
Physically, this means that
correlations decay exponentially as a function of distance, and
that the system has a unique Gibbs measure and no phase
transition.

\dima{The following recent result of Dyer, Sinclair, Vigoda and
Weitz~\cite{dsvw} 
(see also the lecture notes by
Martinelli~\cite{martinelli}) 
relates optimal temporal mixing with
spatial mixing:} \dima{if the boundary} constraints are {\em
permissive}, i.e., if a finite region can always be colored no
matter how we color its boundary, \dima{and if the} heat-bath
dynamics on some finite block mixes in $O(n \log n)$ time, then
the system has strong spatial mixing.  As they point out,
$q$-colorings are permissive for any $q \ge \D+1$.
\mik{Therefore, Theorem 1, which states that $M(2,3)$ has optimal temporal mixing,
implies the following result about spatial correlations.}
\begin{corollary}
\label{cor:spatial} The uniform measure on the set of
$q$-colorings of the square lattice, or equivalently the
zero-temperature antiferromagnetic $q$-state Potts model on the
square lattice, has strong spatial mixing for $q \ge 6$.
\end{corollary}

\dima{As mentioned above, physicists conjecture spatial mixing
for $q \ge 4$. In the last section we discuss to what extent our
techniques might be extended to $q=4,5$.}

\subsection{Markov Chains, Mixing Times and \dima{Earlier Work}}

Given a Markov chain $M$, let $\pi$ be its stationary distribution and
$P_x^t$ be the probability distribution after $t$ steps starting with an
initial \cris{configuration} $x$.  Then, for a given $\eps > 0$, the {\em $\eps$-mixing
time} of $M$ is
\[ \tau_\eps = \max_x \min \left\{ t : \;
   \left\| P_x^t - \pi \right\| < \eps \right\}
\]
where $\left\| P_x^t - \pi \right\|$ denotes the total variation distance
\[ \left\| P_x^t - \pi \right\|
   = \frac{1}{2} \sum_y \left| P_x^t(y) - \pi(y) \right|
\enspace .
\]
In this paper we will often adopt the common practice of suppressing
the dependence on $\eps$, which
is typically logarithmic, and speak just of the mixing time $\tau$ for
fixed small $\eps$. Thus the mixing time becomes a function of $n$, the number
of vertices, alone.
We say that a Markov chain has {\em rapid
mixing} if $\tau = \poly(n)$, and {\em optimal (temporal) mixing}
if $\tau = O(n \log n)$.

The most common Markov chain for \cris{the Potts model is} {\em Glauber
dynamics}.  There are several variants of this in the literature,
but for colorings we fix the following definition, \cris{which applies at zero temperature.}
At each step, choose a random vertex $v \in G$.  Let $S$ be the set of colors,
and let $T$ be the set of colors taken by $v$'s neighbors.  Then
choose a color $c$ uniformly at random from $S \setminus T$,
i.e., from among the colors consistent with the coloring of
$G-\{v\}$, and recolor $v$ with $c$.  

Independently, Jerrum~\cite{j} and Salas and Sokal~\cite{salassokal} 
proved that for $q$-colorings on a graph of maximum degree $\D$ the
Glauber dynamics \mik{(i) is ergodic for $q \ge \D+2$ 
(this holds for fixed boundary conditions as well)
and (ii)} has optimal mixing for $q > 2\D$. \mik{For $q=2\D$, Bubley
and Dyer~\cite{bd} showed that it mixes in $O(n^3)$ time and Molloy~\cite{mm}
showed that it has optimal mixing.}   
Since the square lattice has $\D = 4$, these results imply optimal mixing for 
$q \ge 8$.

Dyer and Greenhill~\cite{dg} considered a ``heat bath'' Markov chain
which updates both endpoints of a random edge simultaneously, and showed
that it has optimal mixing for $q \ge 2\D$.  By widening
the updated region to include a vertex and all of its neighbors, Bubley,
Dyer and Greenhill~\cite{bdg} showed optimal mixing for 
$q \ge 7$ for triangle-free graphs \mik{with maximum degree 4}, 
\cris{which includes} the square lattice.  

Another commonly used Markov chain is the Kempe chain
algorithm, known in physics as the zero-temperature case of the
{\em Wang-Swendsen-Koteck\'y} algorithm~\cite{wsk1,wsk2}. It
works as follows: we choose a random vertex $v$ and
a color $b$ which differs from $v$'s current color $a$.
We construct the largest connected subgraph containing $v$ which is
colored with $a$ and $b$, and recolor this subgraph by switching
$a$ and $b$.  In a major breakthrough, Vigoda~\cite{ev} showed that
a similar Markov chain has optimal mixing for $q > (11/6)\D$,
and \mik{that} this implied that the Glauber dynamics and the Kempe chain algorithm
both have rapid mixing for $q\ge (11/6)\D$.
However, for the square lattice this again gives only $q \ge 8$.

For $q=3$ on the square lattice, Luby, Randall and
Sinclair~\cite{lrs} showed that a Markov chain including ``tower
moves'' has rapid mixing for any finite simply-connected region
with fixed boundary conditions, and Randall and
Tetali~\cite{randalltetali} showed that this implies rapid mixing
for the Glauber dynamics as well. Recently Goldberg, Martin and
Paterson~\cite{gmp} proved rapid mixing for the Glauber dynamics
on rectangular regions with free boundary conditions, i.e., with
no fixed coloring of the vertices on their boundary. \cris{Unfortunately,} 
the technique of~\cite{lrs,gmp} relies on a bijection between
3-colorings and random surfaces through a ``height
representation'' which does not hold for other values of $q$.

\section{Coupling}

We consider two parallel runs of our Markov chain, $M(2,3)$, with initial colorings $X_0,Y_0$.
We will couple the steps of these chains in such a way that
(i) each chain runs according to the correct distribution on its choices and
(ii) with high probability, $X_t=Y_t$ for some $t=O(n\log n)$.  A now standard fact
in this area is that this implies that the chain mixes in time $O(n\log n)$, i.e.
this implies Theorem~\ref{t1};  this fact was first proved by Aldous \cite{da}
(see also \cite{dgs}).

Bubley and Dyer \cite{bd} introduced the very useful technique of Path Coupling,
via which it suffices to do the following:  Consider any two
colorings $X, Y$ which 
\cris{have a Hamming distance one, i.e., which} differ on exactly one vertex, 
and carry out a single step
of the chain on $X$ and on $Y$, producing two new colorings $X', Y'$.
We will prove that we can couple these two steps such that
(i) each step is selected according to the correct distribution,
and (ii) the expected \cris{Hamming distance} between $X'$ and $Y'$ 
is at most $1-\e/n$ for some constant $\e>0$.  \cris{Thus the expected change 
in the Hamming distance between the two colorings is negative, $-\e/n$.}
See, e.g., \cite{dgs} for the formal (by now standard)
details as to why this suffices to prove Theorem~\ref{t1}.

We perform the required coupling as follows. \mik{Let $X$ and $Y$ be two
arbitrary 6-colorings which only disagree at one vertex.} We pick a uniformly random $2 \times 3$
block $S$, and let $C_X$ and $C_Y$ denote the set of permissible recolorings of
 $S$ according to $X,Y$ respectively.  For each $c \in C_X$, we define a carefully
chosen probability distribution $p_c$ on the colorings of $C_Y$.
We pick a uniformly random coloring $c_1 \in C_X$ and in $X$ we recolor
$S$ with $c_1$  to produce $X'$.  We then pick a random coloring $c_2 \in C_Y$
according to the distribution $p_{c_1}$ and in $Y$ we recolor $S$ with $c_2$  to produce $Y'$.
Trivially, \cris{the marginal distribution $\langle S, c_1 \rangle$ is uniform on $S$ 
and $C_X$.} In order to ensure that the same is true of $\langle S, c_2 \rangle$, 
we must have the following property for the set of distributions $\{p_c : c \in C_X\}$:

\begin{equation}\label{c1}
\mbox{for each $c_2\in C_Y$}, \qquad\inv{|C_X|}\sum_{c\in C_X}p_c(c_2)=\inv{|C_Y|} \enspace .
\end{equation}

Suppose that $v$ is the vertex on which $X,Y$ differ.  If $v\in S$ then
$C_X=C_Y$, so we can simply define $c_2=c_1$ (i.e., $p_c(c)=1$ for each $c$)
and this ensures that $X'=Y'$.  If $S$ does not contain $v$ or any neighbor
of $v$, then again $C_X=C_Y$ and by defining $c_2=c_1$ we ensure that
$X',Y'$ differ only on $v$.  If $v$ is not in $S$ but is adjacent to
a vertex in $S$, then $C_X \neq C_Y$ and so, 
\cris{depending on our coupling,}
it is quite possible that $c_2\neq c_1$ and so $X',Y'$ will differ on
one or more vertices of $S$ as well as on $v$.

For any pair $C_X, C_Y$, we let $H(C_X, C_Y)$ denote the expected number
of vertices in $S$ on which $c_1,c_2$ differ.  For every possible pair
$C_X, C_Y$ we obtain a coupling satisfying (\ref{c1}) and:
\begin{equation}\label{c2}
H(C_X, C_Y)<0.52 \enspace .
\end{equation}
\cris{Thus the expected change in the Hamming distance between the two colorings is
$-1$ if $v \in S$ and less than $0.52$ if $v$ is adjacent to $S$.}

\medskip
\noindent {\bf Proof of Theorem \ref{t1}.}   As described above,
it suffices to prove that for any choice of $X,Y$ differing only
at $v$, the expected \cris{Hamming distance between $X'$ and $Y'$} is less
than $1-\e/n$ for some $\e>0$, or equivalently that the expected change 
in the Hamming distance is less than $-\e/n$.

\remove{
the probability that $S$ contains $v$ is $|S|/n=6/n$, and for any such choice of $S$, $X'=Y'$.
The probability that $S$ contains a neighbor of $v$ but does not contain $v$
is easily seen to be \cris{at most} $10/n$; for any such choice of $S$, the expected
number of vertices on which $X',Y'$ differ is less than $1.52$.  Therefore,
the overall expected number of vertices on which $X',Y'$ differ is at most
}

\cris{Let us consider toroidal boundary conditions first; there are $n$ possible blocks 
from which the algorithm will choose, one for each vertex.  Each vertex is contained 
in $6$ blocks and is adjacent to $10$ blocks, so $v \in S$, or $v$ is adjacent to $S$, 
with probability $6/n$ or $10/n$ respectively.  Thus the expected change in the 
Hamming distance is less than
\[ -1\times \frac{6}{n} + 0.52 \times \frac{10}{n} = -\frac{0.8}{n} \enspace . \]
}

\remove{
\cris{For finite lattices of height $h$ and width $w$, the number of possible blocks 
\is $n-2h-w+2$; let us denote this $m$.  There are several cases we have to deal with, 
depending on $v$'s location: in each one, let $A$ be the number of blocks containing $v$, 
and let $B$ be the number of blocks to which $v$ is adjacent.  Then the expected change
in the Hamming distance is less than 
\[ -1 \times \frac{A}{m} + 0.52 \times \frac{B}{m} = \frac{0.52 B - A}{m} \]
We need to ensure that this is negative in each case, or equivalently, 
that $A/B > 0.52$.  

For vertices deep in the interior of the lattice, 
we have $A=6$ and $B=10$ as in the toroidal case.  
For other types of vertices, we have the following values of $A$ and upper bounds on $B$: 
\[
\begin{array}{ccc}
\mbox{type} & A & B \le \\ \hline
\mbox{corner} & 1 & 2 {\bf argh} \\
\mbox{top or bottom edge, adjacent to corner} & 2 & 3 \\
\mbox{top or bottom edge, not adjacent to corner} & 3 & 5 \\ 
\mbox{left or right edge} & 2 & 4 {\bf argh} 
\end{array}
\]
{\bf We really have a problem here; unless I am mistaken, our argument fails for 
e.g. corners and several other types of vertices too.  How did we miss this before?}
}
\proofend
}
For \cris{rectangular regions with free boundary conditions} 
the situation is a little more complicated.
\remove{; the Markov chain 
we have defined will work without modification on the torus, but requires some 
extra instructions to deal \cris{with finite regions}.}
A natural extension of 
the basic procedure is to chose any block that lies entirely in the region. 
This, however, means that a region of height $h$ and width $w$ with $n$ vertices 
will only have $n-2h-w+2$ possible $2 \times 3$ blocks; furthermore, when the distinguished 
vertex $v$ is on or close to the boundary, the number of blocks containing $v$  
is less than the 6 we have when $v$ is snugly in the interior, thereby lowering 
our chances of decreasing the Hamming distance.

It turns out that if we are dealing with free boundary conditions the natural 
extension will work anyway, since then the maximum expected change in the Hamming 
distance when $v$ is on or close to the boundary and {\em adjacent} to our block 
is low enough to counteract the reduced chance that our choice of block will 
contain $v$. In particular, we obtained these values for the expected change in 
the Hamming distance when $v$ is adjacent to the block:
\begin{center}
\begin{tabular}{llcr}
$v$ is on boundary, block is not: & $< 0.52$ & & {\em i.e., the interior case} \\
block is on boundary, $v$ is not: & $< 0.503$ & & \\
both $v$ and block are on boundary: & $< 0.390$ & &
\end{tabular}
\end{center}
As an example, when $v$ is in the corner of the lattice we have one choice of a
block containing $v$, and two choices of a block adjacent to $v$; both of the 
adjacent blocks are also on the boundary too, so the expected change in 
the Hamming distance is less than
\[ -1\times \frac{1}{n-2h-w+2} + 0.39 \times \frac{2}{n-2h-w+2} = 
-\frac{0.22}{n-2h-w+2} \enspace . \]
It can be easily verified that in the other cases where $v$ is on or close to a 
free boundary the values given above are also well within what we need for rapid 
mixing.

\cris{If we have to deal with fixed, arbitrary boundary conditions---that is, 
a fixed proper coloring of the sites just outside the region---applying our 
Markov process directly will not give us the same result. The problem is 
that the ratio between the number of blocks containing $v$ and adjacent to $v$ 
is the same as in the free boundary case, but the expected change in Hamming 
distance for any block adjacent to $v$ is the same as in the toroidal case, 
as opposed to the lower values we obtain for free boundary conditions above.}

An alternative modification of our basic Markov process that will work is to 
choose any block that contains {\em any} vertex in the region, and then 
(properly) recolor all the vertices in that block which are also in the region.  
For a lattice of height $h$ and width $w$ with $n$ vertices this gives us 
$n+2h+w+2$ possible $2 \times 3$ blocks to chose from; more importantly, 
for every vertex in the lattice there are 6 available blocks which contain it.

The expected change in the Hamming distance when the block is on the boundary but 
fully contained in the lattice is now no different than the expected change for a 
block on the interior, but we do have to deal with cases where the block may be 
partially outside the lattice. When this happens we can merely treat 
the sub-block that we do recolor in exactly the same manner as we would treat a 
smaller block being recolored on the interior (that is, we reduce the size of  
the block to contain only the recolorable part). The calculations for smaller 
blocks adjacent to $v$ were of course already done in order to confirm previous 
results or rule out the possibility of rapid mixing for block sizes below 
$2 \times 3$, and the results were:
\begin{center}
\begin{tabular}{ccccr}
Sub-block size & $\quad$ & Max expected change & & \\ \hline
$1 \times 1$ & & $\leq 0.50$ & & \\
$2 \times 1$ or $1 \times 2$ & & $< 0.524$ & & \\
$1 \times 3$ & & $< 0.514$ & & \\
$2 \times 2$ & & $< 0.508$ & & \\
$2 \times 3$ & & $< 0.52$ & & {\em (i.e. the interior case)}
\end{tabular}
\end{center}
As an example, when $v$ is in the corner of the lattice we now have 6 choices 
of a block containing $v$, and five choices of a block adjacent to $v$; 2 of 
these choices are full $2 \times 3$ blocks, and the remaining choices have 
sub-block sizes of $2 \times 1$, $2 \times 2$, and $1 \times 3$ respectively. 
Hence the expected change in the Hamming distance is less than
\[ -1\times \frac{6}{n+2h+w+2} + \frac{0.52 \times 2 + 0.524 + 0.508 + 
0.514}{n+2h+w+2} = -\frac{3.414}{n+2h+w+2} \enspace . \]
We could work out the cost explicitly for the rest of the cases where the 
distinguished vertex $v$ is on or near the boundary, but it is easier just to 
note that in no case can the expected cost of coloring a block adjacent to $v$ be 
more than $0.524$, and there can never be more than 10 adjacent blocks to choose 
from, so the expected change in the Hamming distance can never be more than 
\[ -\frac{0.76}{n+2h+w+2} \enspace . \]
\proofend
\vspace{0.5cm}

Of course, we still need to prove that the desired couplings exist for
each possible $C_X,C_Y$.  These couplings were found with the aid of computer programs.
In principle, for any pair $C_X,C_Y$, searching for the coupling that minimizes
$H(C_X,C_Y)$ subject to (\ref{c1}) is simply a matter of solving a linear program
and so can be done in polynomial time.  However, the number of variables is
$|C_X| |C_Y|$ which, {\em a priori}, can be roughly $(5^6)^2$.  
Furthermore, the number of possible pairs $X,Y$ is roughly $6^{10}$, and even
after eliminating pairs which are redundant by symmetry, it is enormous.
To \cris{deal with this combinatorial explosion} 
we designed a fast heuristic which, rather than finding the best coupling
for a particular pair, simply finds a very good coupling; i.e., one that
satisfies (\ref{c2}).  The code used can be found at
\verb+www.cs.toronto.edu/~fvb+.  We provide a more detailed
description in the next section.

\section{The Programs Used}

{\bf Method of the computation:} Let $R$ denote the {\em rim} vertices,
that is, those vertices which are adjacent to but outside of
the block $S$. We call a 
coloring of the vertices of $R$ a {\em rim coloring}.
For each possible
pair of rim colorings $X,Y$ which differ only at a vertex $v\in R$,
we need to find a coupling between the extensions $C_X$ and $C_Y$ of
$X,Y$ to $S$, so that the couplings satisfy (\ref{c1}) and (\ref{c2}).
These couplings were found with a small suite of programs working in two
phases. In the first phase, exhaustive lists of pairs of rim colorings (reduced
by equivalence with respect to allowable block colorings) were
generated. In the second phase, for each pair $X,Y$,
we generated $C_X, C_Y$ separately; these were then coupled, satisfying (\ref{c1}),
in a \cris{nearly} optimal way to
obtain a bound on $H(C_X,C_Y)$ that satisfies (\ref{c2}).

{\bf Implementation:} All programs take the following parameters:
number of colors, block dimensions, and an integer denoting the position
of the distinguished vertex $v$ with respect to the block (0 if adjacent
to the corner, $+i$ if adjacent to the $i$th vertex along the top of
the block, and $-i$ if adjacent to the $i$th vertex along the side of
the block). \mik{We assume that $v$ has color 0 in $X$ and 1 in $Y$.
Thus,} if one specifies a rim coloring $X$ and
the position of $v$, then this determines $Y$. 
For each coloring $X$ we determined a good
coupling for each non-equivalent position of $v$.

By default the programs generate rim colorings and couplings on the
assumption that the block is not on the boundary of the lattice
(i.e., all rim vertices potentially constrain the allowable block colorings). 
\remove{
Boundary cases, however, can easily be simulated by using
values in the rim colorings that are outside the range determined by
the number-of-colors parameter. Blocks on the boundary were only checked
when the analysis of the non-boundary blocks yielded promising values;
in all cases we found that the maximum cost for boundary blocks was lower than for the
associated non-boundary blocks.
}
Free boundary cases, however, can easily be simulated by using values in the 
rim colorings that are outside the range determined by the number-of-colors 
parameter. These were only checked when the analysis of the non-boundary blocks 
yielded promising values. As mentioned above, the fixed arbitrary boundary cases 
required no new \remove{per block expected change} calculations, since from the block's 
``point of view'' there is no difference (in the worst case) between being on an 
arbitrarily colored boundary and being in the interior.  Thus, we simply  
reused previously calculated values for smaller blocks on the interior to verify 
that the modified Markov process would work.


{\bf Generating rim colorings:} Since the calculations required for
phase 2 were much more time-consuming than those for phase 1, the rim
coloring generation procedure was designed to minimize the number of
colorings output rather than the time used generating them. A rim
coloring is represented by a vector of colors used on the rim,
starting from the distinguished vertex $v$ and going clockwise around
the block.  \cris{If the set of colors is $\{0,\ldots,5\}$,} 
we can assume by symmetry that $v$'s color 
is 0 in $X$ and 1 in $Y$.
\remove{, $0|1$ is always the first element (0 is
used in the actual output, 1 is understood).}
The following reductions
were applied to avoid equivalent rim colorings: reduction by color
isomorphism (colors 2 and above), by exchange of colors 0 and 1, by
exchange of colors of vertices adjacent to the corners of the block,
and by application of flip symmetries where applicable.

{\bf Finding a coupling for particular rim colorings:} Two
programs were used for each rim
coloring $X$, and position $i$ of $v$.
In each, the initial operation is the generation of all
compatible lattice colorings; this is
done separately for col$(v)=0$ and col$(v)=1$
(i.e., for $C_X$ and $C_Y$). The first program
creates a set of linear programming constraints that is readable by the
program {\tt lp-solve} (by Michel Berkelaar of the Eindhoven University
of Technology; it is available with some Linux distributions). As
mentioned above, time and space requirements made use of this procedure
feasible only for checking individual rim colorings, and even then the
block size had to be fairly modest. 

The second program calculates an
upper bound on the optimal cost using a greedy algorithm to create a
candidate coupling. Given sets of colorings $C_X$ and $C_Y$ of size
$m_X$ and $m_Y$ respectively, the algorithm starts by assigning
``unused'' probabilities \mik{of $1/m_X$ and $1/m_Y$ respectively} to the
individual colorings. Then, for each distance \mik{$d=0,1,...,6$}, for each
coloring $c_1$ in $C_X$ it traverses $C_Y$ looking for a coloring
$c_2$ which differs from $c_1$ on exactly $d$ vertices.
When such an $c_2$ is found it removes
the coloring with the lower unused probability $p$ from its list and
reduces the unused probability $p'$ of the other to $p'-p$; the distance
$d \cdot p$ is added to the total distance so far. The order in which
the lists of colorings $C_X$ and $C_Y$ is traversed does affect the
solution, so an optional argument is available that allows the user to
select one of several alternatives.

This heuristic does not guarantee an optimal solution, and
with some blocks and particular rim colorings the coupling it generates
is far from the best. However, for the rim colorings we are most
interested in (ones where $H(C_X,C_Y)$ is high for all couplings) it seems to
consistently give results that are optimal or very close (within 2\%).
We cannot give a rigorous bound on the running time, but a cursory
analysis and empirical evidence suggest that it runs in roughly
${\mathcal O}(m \log{m})$ time, where $m$ is the number of
compatible block colorings. Because the heuristic is so much faster than
the LP solver, our general procedure was as follows: (1) Use the
heuristic with the default traversal order to calculate bounds on the
expected distance for all the rim colorings generated in phase one.
(2) When feasible, use the LP solver on those rim colorings that had
the highest value of $H(C_X,C_Y)$, to obtain an exact value for their maximum.
(3) For larger blocks or more colors than could be comfortably handled
by the LP solver, use all available traversal orders on those
rim colorings that had the maximum value of $H(C_X,C_Y)$
to obtain as tight a bound as possible within a feasible time.

{\bf Results of the computations:} Computations were run on various block 
sizes and numbers of colors in order to check the correctness of the
programs, and \mik{also to} collect data which could be used to estimate
running times and maximum expected distance for larger block 
dimensions. For 7- and 8-colorings our results corresponded
well with previous work on the problem (e.g.~\cite{bd}).

For 6-colorings, we checked $1 \times k$ blocks for $k \leq 5$, as
well as $2 \times 2$ and $2 \times 3$ blocks.  For all but the last of
these the maximum expected distance we obtained was too large to give us
rapid mixing. 
\cris{The $2 \times 3$ subgrid has 2 non-equivalent positions
with respect to the rim: the corner (to which 8 rim vertices are
adjacent) and the middle of the top or bottom side (to which 2 rim vertices are
adjacent).  Denote these positions $1$ and $2$ respectively.}
For each $X,Y$ with $v$ in position 1, we obtained a coupling satisfying:
\[H(C_X,C_Y)\leq 0.5118309760 \enspace . \]
For each $X,Y$ with $v$ in position 2, we obtained a coupling satisfying:
\[H(C_X,C_Y)\leq 0.4837863092 \enspace . \]
Thus, in each case we satisfy (\ref{c2}) as required.

{\bf A slightly stronger output:} By examining the problem a bit more closely,
we see that condition (\ref{c2}) is sufficient, but not necessary, for our
purposes.  Let $H_i$ denote the maximum of $H(C_X,C_Y)$ over the couplings found
for all pairs $X,Y$ where $v$ is in position $i$, and let $\mult_i$ denote
the number of rim vertices adjacent to position $i$.  Then, being more careful
about the calculation used in the proof of Theorem \ref{t1}, and extending it
to a general $a\times b$ block, we see that
the overall expected change in the Hamming distance is at most 
(in, say, the toroidal case where the number of possible blocks is $n$)
\[
-1 \times \frac{ab}{n} + \sum_i \mult_i \times \frac{H_i}{n} \enspace , \]
a smaller value than that used in the proof of Theorem 1,
where we (implicitly) used $(\max_i H_i)\times\sum_i \mult_i$ rather
than $\sum_i \mult_i\times H_i$.
Our programs actually compute this smaller value.  Even so,
we could not obtain suitable couplings for any block size smaller
than $2\times 3$.


\section{Rapid Mixing: Glauber and Kempe Chain Dynamics}

In this section we prove Theorem~\ref{t2}, showing rapid mixing for
the Glauber and Kempe chain dynamics, by following the techniques and
presentation of Randall and Tetali~\cite{randalltetali}.

\remove{
Suppose $Q$ is a Markov chain whose mixing time we would like to
bound, and $\tQ$ is another Markov chain for which we already
have a bound.  Let $E$ and $\tE$ and denote the edges of these Markov
chains, i.e., the pairs $(x,y)$ such that the transition probabilities
$Q(x,y)$ and $\tQ(x,y)$, respectively, are positive.  Now, for each
edge of $\tQ$, i.e., each $(x,y) \in \tE$, choose a fixed path
$\gamma_{x,y}$ using the edges of $Q$: that is, choose a series of
states $x=x_1, x_2, \ldots, x_k=y$ such that $(x_i,x_{i+1}) \in E$ for
$1 \le i < k$.  Denote the length of such a path $|\gamma_{x,y}|=k$.
Furthermore, for each $(z,w) \in E$, let $\Gamma(z,w) \subseteq \tE$
denote the set of pairs $(x,y)$ such that $\gamma_{x,y}$ uses the edge
$(z,w)$.  Finally, let
\[ A = \max_{(z,w) \in E}
   \left[ \frac{1}{Q(z,w)}
          \sum_{(x,y) \in \Gamma(z,w)} |\gamma_{x,y}| \,\tQ(x,y)
   \right]
\enspace .
\]
Note that $A$ depends on our choice of paths.

By combining bounds on the mixing time in terms of the spectral
gap~\cite{diaconisstroock,sinclairjerrum,sinclair} with an upper bound
on $Q$'s spectral gap in terms of $\tQ$'s due to Diaconis and
Saloff-Coste~\cite{ds-c}, we obtain the following upper bound on $Q$'s
mixing time:
\begin{theorem}
\label{thm:comp}
Let $Q$ and $\tQ$ be reversible Markov chains on $q$-colorings of a graph of $n$
vertices whose unique stationary distribution is the uniform
distribution.  Let $\tlambda_1$ be the largest eigenvalue of $\tQ$'s
transition matrix smaller than $1$, let $\tau_\eps$ and $\ttau_\eps$
denote the $\eps$-mixing time of $Q$ and $\tQ$ respectively, and
define $A$ as above.  Then for any $\eps \le 1/4$,
\[ \tau_\eps \le \frac{4 \log q}{\tlambda_1} A n \ttau_\eps \enspace . \]
\end{theorem}
We omit the proof.
The reason for the additional factor of $n$ is the fact that the upper
and lower bounds on the mixing time in terms of the spectral gap are 
$\log \pi(x)^{-1}$ apart, where $\pi$ is the uniform distribution.  Since
there are at most $q^n$ colorings, we have $\log \pi(x)^{-1} \le n
\log q$.  On the square lattice, it is easy to see that there are an
exponentially large number of $q$-colorings for $q \ge 3$, so removing
this factor of $n$ would require a different comparison technique.

Now suppose that $\tQ$ is the block dynamics and $Q$ is the
Glauber or Kempe chain dynamics.  We wish to prove Theorem~\ref{t2}
by showing that $\tau_\eps = O(n \ttau_\eps)$.
By adding self-loops with probability greater than $1/2$ to the block
dynamics, we can ensure that the eigenvalues of $\tQ$ are positive
with only a constant increase in the mixing time.  Therefore, it suffices
to find a choice of paths for which $A$ is constant.  Since, for all
three of these Markov chains, each move occurs with probability
$\Theta(1/n)$, if $|\gamma_{x,y}|$ and
$|\Gamma(z,w)|$ are constant then so is $A$.
}
Suppose $P$ and $\tP$ are two Markov chains on the same state space
with the same stationary distribution $\pi$, and that we already have a bound on 
the mixing time of $\tP$ while we would like to obtain a bound for that of $P$.
Let $E(P)$ and $E(\tP)$ denote the edges of these Markov chains, i.e., the pairs 
$(x,y)$ such that the transition probabilities $P(x,y)$ and $\tP(x,y)$, 
respectively, are positive.  Now, for each edge of $\tP$, i.e., each
$(x,y) \in E(\tP)$, choose a fixed path $\gamma_{x,y}$ using the edges of $P$: 
that is, choose a series of states $x=x_0, x_1, x_2, \ldots, x_k=y$ such that 
$(x_i,x_{i+1}) \in E(P)$ for $0 \le i < k$.  Denote the length of such a path 
$|\gamma_{x,y}|$. Furthermore, for each $(z,w) \in E(P)$, let 
$\Gamma(z,w) \subseteq E(\tP)$ denote the set of pairs $(x,y)$ such that 
$\gamma_{x,y}$ uses the edge $(z,w)$.  Finally, let
\[ A = \max_{(z,w) \in E(P)}
   \left[ \frac{1}{\pi(z)P(z,w)}
          \sum_{(x,y) \in \Gamma(z,w)} |\gamma_{x,y}| \, \pi(x) \, \tP(x,y)
   \right]
\enspace .
\]
Note that $A$ depends on our choice of paths.

By combining bounds on the mixing time in terms of the spectral
gap~\cite{diaconisstroock,sinclairjerrum,sinclair} with an upper bound
on $P$'s spectral gap in terms of $\tP$'s due to Diaconis and
Saloff-Coste~\cite{ds-c}, we obtain the following upper bound on $P$'s
mixing time:
\begin{theorem}
\label{thm:comp}
Let $P$ and $\tP$ be reversible Markov chains on $q$-colorings of a graph of $n$
vertices whose unique stationary distribution is the uniform
distribution.  Let $\lambda_1(\tP)$ be the largest eigenvalue of $\tP$'s
transition matrix smaller than $1$, let $\tau_\eps$ and $\ttau_\eps$
denote the $\eps$-mixing time of $P$ and $\tP$ respectively, and
define $A$ as above.  Then for any $\eps \le 1/4$,
\[ \tau_\eps \le \frac{4 \log q}{\lambda_1(\tP)} A n \ttau_\eps \enspace . \]
\end{theorem}
We omit the proof.
The reason for the additional factor of $n$ is the fact that the upper
and lower bounds on mixing time in terms of the spectral gap are
$\log 1/\pi_*$ apart, where $\pi_*$ is the minimum of $\pi(x)$ taken over all 
states $x$. Since $\pi$ in this case is the uniform distribution
and there are at most $q^n$ colorings, we have
$\log 1/\pi(x) \le n \log q$.  On the square lattice, it is easy to see that 
there are an exponentially large number of $q$-colorings for $q \ge 3$, so 
removing this factor of $n$ would require a different comparison technique.

Now suppose that $\tP$ is the block dynamics and $P$ is the
Glauber or Kempe chain dynamics.  We wish to prove Theorem~\ref{t2}
by showing that $\tau_\eps = O(n \ttau_\eps)$.
By adding self-loops with probability greater than $1/2$ to the block
dynamics, we can ensure that the eigenvalues of $\tP$ are positive
with only a constant increase in the mixing time.  Therefore, it suffices
to find a choice of paths for which $A$ is constant.  Since, for all
three of these Markov chains, each move occurs with probability
$\Theta(1/n)$, if $|\gamma_{x,y}|$ and
$|\Gamma(z,w)|$ are constant then so is $A$.

In fact, for $q \ge \D + 2$, we can carry out a block move on any
finite neighborhood with Glauber moves.  We need to flip each vertex $u$
in the block to its new color; however, $u$'s flip is blocked by a
neighbor $v$ if $v$'s current color equals $u$'s new color.  Therefore,
we first prepare for $u$'s flip by changing $v$ to a color which differs
from $u$'s new color as well as that of $v$'s $\D$ neighbors.  If
the neighborhood has $m$ vertices, this gives $|\gamma_{x,y}| \le
m(\D+1)$, or $|\gamma_{x,y}| \le 30$ for $M(2,3)$.
(With a little work we can reduce this to $13$.)

For the Kempe chain dynamics, recall that each move of the chain
chooses a vertex $v$ and a color $b$ other than $v$'s current color.
If $b$ is the color which the Glauber dynamics would assign to $v$,
then none of $v$'s neighbors are colored with $b$, and the Kempe
chain move is identical to the Glauber move.  Since this happens with
probability $1/q$, the above argument applies to Kempe chain moves as
well, and again we have $|\gamma_{x,y}| \le m(\D+1)$.
Moreover, we only need to consider moves that
use  Kempe chains of size $1$.

Finally, since each vertex appears in only $m=6$ blocks, the number of
block moves that use a given Glauber move or a given Kempe chain move
of size $1$ is bounded above by $m$ times the number of pairs of colorings
of the block.  Thus $|\Gamma(z,w)| \le m (q^m)^2$, and we are done.

An interesting open question is whether we can prove {\em optimal}
temporal mixing for the Glauber or Kempe chain dynamics.  One possibility
is to use log-Sobolev inequalities as in~\cite{cesi}.  We leave this as a
direction for further work.

\section{Conclusion: \cris{Larger Blocks and Smaller $q$?}}

We have run our programs on $2 \times 4$ and $3 \times 3$ blocks to
see if we could achieve rapid mixing on 5 colors, but in both cases
the largest values of $H(C_X,C_Y)$ were too high. It 
may be
that
rapid mixing on 5 colors is possible by recoloring a $3 \times 4$
block, based on the decrease of the ratio of $\max H(C_X,C_Y) \cdot |R|$
to $|S|$ as the dimensions increase; similar reasoning leads us to
believe that rapid mixing using a $2 \times k$ block is possible,
but we would probably need a $2 \times 10$ block or larger to achieve
success.
Unfortunately, doing the calculations for $3 \times 4$ blocks is a daunting
proposition. The problem is exponential in two directions at once
(number of rim colorings, and number of block colorings for each rim
coloring), \cris{so this would require a huge increase in the running time.}

\bigskip
{\bf Acknowledgments.}  We are grateful to Leslie Goldberg,
Dana Randall and Eric
Vigoda for helpful discussions, and to an anonymous referee for
pointing out a technical error.
Recently, L. Goldberg, R. Martin and M. Paterson have reported
obtaining the main result of this paper independently.


\begin{thebibliography}{99}

\bibitem{da}
Aldous, D.:
Random walks on finite groups and rapidly mixing Markov chains.
S\'eminaire de Probabilit\'es XVII 1981/82 (Dold, A. and Eckmann, B., eds.), Springer Lecture Notes in Mathematics {\bf 986} (1986) 243--297

\bibitem{bd}
Bubley, R., Dyer, M.:
Path coupling: a technique for proving rapid mixing in Markov chains.
Proc.\ 28th Ann.\ Symp.\ on Found. of Comp. Sci. (1997) 223--231

\bibitem{bdg}
Bubley, R., Dyer, M., Greenhill, C.:
Beating the $2 \D$ bound for approximately counting colourings: A computer-assisted proof of rapid mixing.
Proc.\ 9th Ann.\ ACM-SIAM Symposium on Discrete Algorithms (1998) 355--363

\bibitem{bdgj}
Bubley, R., Dyer, M., Greenhill, C., Jerrum, M.:
On approximately counting colourings of small degree graphs.
SIAM J. Comp. {\bf 29} (1999) 387--400

\bibitem{cesi}
Cesi, F.:
Quasi-factorization of the entropy and logarithmic Sobolev inequalities for Gibbs random fields.
Probability Theory and Related Fields {\bf 120} (2001) 569--584

\bibitem{ds-c}
Diaconis, P., Saloff-Coste, L.:
Comparison theorems for reversible Markov chains.
Annals of Applied Probability {\bf 6} (1996) 696--730

\bibitem{diaconisstroock}
Diaconis, P., Stroock, D.:
Geometric bounds for eigenvalues of Markov chains.
Annals of Applied Probability {\bf 1} (1991) 36--61



\bibitem{dg}
Dyer, M., Greenhill, C.:
A more rapidly mixing Markov chain for graph colorings.
Random Structures and Algorithms {\bf 13} (1998) 285--317

\bibitem{dgs}
Dyer, M., Greenhill, C.:
Random walks on combinatorial objects.
Surveys in Combinatorics, 1999 (Lamb, J. and Preece, D., eds.), Cambridge University Press, J., 1999, 101--136


\bibitem{dsvw}
Dyer, M., Sinclair, A., Vigoda, E., Weitz, D.:
Mixing in time and space for lattice spin systems: a combinatorial view.
Proc. RANDOM (2002) 149--163

\bibitem{ferreirasokal}
Ferreira, S.,  Sokal, A.:
Antiferromagnetic Potts models on the square lattice: a high-precision Monte Carlo study.
J. Statistical Physics {\bf 96} (1999) 461--530

\bibitem{gmp}
Goldberg, L., Martin, R., Paterson, M.:
Random sampling of 3-colourings in $\Z^2$.
Random Structures and Algorithms (to appear)


\bibitem{j}
Jerrum, M.:
A very simple algorithm for estimating the number of $k$-colorings of a low-degree graph.
Random Structures and Algorithms {\bf 7} (1995), 157--165


\bibitem{lrs}
Luby, M., Randall, D., Sinclair, A.:
Markov chain algorithms for planar lattice structures.
SIAM Computing {\bf 31} (2001) 167--192

\bibitem{martinelli}
Martinelli, F.:
Lectures on Glauber dynamics for discrete spin models.
Lectures on Probability Theory and Statistics, Saint-Flour 1997, Springer Lecture Notes in Mathematics {\bf 1717} (1999) 93-191

\bibitem{mm}
Molloy, M.:
Very rapidly mixing Markov Chains for $2\D$-colourings and for independent sets in a 4-regular graph.
Random Structures and Algorithms {\bf 18} (2001) 101--115

\bibitem{moorenewman}
Moore, C., Newman, M.:
Height representation, critical exponents, and ergodicity in the four-state triangular Potts antiferromagnet.
J. Stat. Phys. {\bf 99} (2000) 661--690

\bibitem{vortex}
Moore, C., Nordahl, M., Minar, N., Shalizi, C.:
Vortex dynamics and entropic forces in antiferromagnets and antiferromagnetic Potts models.
Physical Review E {\bf 60} (1999) 5344--5351

\bibitem{randalltetali}
Randall, D., Tetali, P.:
Analyzing Glauber dynamics by comparison of Markov chains.
J. Mathematical Physics {\bf 41} (2000) 1598--1615

\bibitem{salassokal}
Salas, J., Sokal, A.:
Absence of phase transition for antiferromagnetic Potts models via the Dobrushin uniqueness theorem.
J. Statistical Physics {\bf 86} (1997) 551--579

\bibitem{sinclair}
Sinclair, A.:
Algorithms for random generation and counting: a Markov chain approach.
Birkhauser, Boston, 1993, pp. 47--48. 

\bibitem{sinclairjerrum}
Sinclair, A., Jerrum, M.:
Approximate counting, uniform generation, and rapidly mixing Markov chains.
Information and Computation {\bf 82} (1989) 93--133

\bibitem{sokalunsolved}
Sokal, A.:
A personal list of unsolved problems concerning lattice gases and antiferromagnetic Potts models.
Talk presented at the conference on Inhomogeneous Random Systems, Universit\'e de Cergy-Pontoise, January 2000. Markov Processes and Related Fields {\bf 7} (2001) 21--38

\bibitem{ev}
Vigoda, E.:
Improved bounds for sampling colorings.
J. Mathematical Physics {\bf 41} (2000) 1555--1569

\bibitem{wsk1}
Wang, J., Swendsen, R., Koteck\'y, R.:
Physical Review Letters {\bf 63} (1989), 109--112

\bibitem{wsk2}
Wang, J., Swendsen, R., Koteck\'y, R.:
Physical Review B {\bf 42} (1990), 2465--2474

\end{thebibliography}
\end{document}